\documentclass[%
 reprint,
nofootinbib,
 amsmath,amssymb,
 aps, amsfonts,
prd,
floatfix
]{revtex4-2}
\usepackage{microtype}

\usepackage{hyperref}
\hypersetup{
    colorlinks=true,
    linkcolor=[rgb]{0,0,0.6},      
    urlcolor=[rgb]{0,0,0.6},
    citecolor=[rgb]{0,0,0.6}
    }

\usepackage{subcaption}
\usepackage[normalem]{ulem}
\usepackage{soul}
\setcitestyle{numbers,open={[},close={]}}
\usepackage{graphicx}
\usepackage{dcolumn}
\usepackage{bm}

\usepackage[super]{nth}
\usepackage{booktabs}
\usepackage[version=4]{mhchem}
\usepackage{mathtools} 
\usepackage{float}

\DeclarePairedDelimiter\autobracket{(}{)}
\newcommand{\pr}[1]{\autobracket*{#1}}

\usepackage{Zallman,lettrine}

\newcommand{\dd}{\mathrm{d}}

\newcommand{\ket}[1]{\left|#1\right>}

\newcommand{\orcid}[1]{\href{https://orcid.org/#1}{\includegraphics[width=10pt]{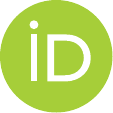}}}

\begin{document}

\preprint{APS/123-QED}

\title{Neutrino Flavor Transformation in Collapsing Supermassive Objects}

\author{Kyle S. Kehrer\,\orcid{0000-0002-8714-1599}}
\email{kkehrer@ucsd.edu}
\affiliation{Department of Physics, University of California San Diego, La Jolla, California 92093, USA}
\author{George M. Fuller\,\orcid{0000-0002-4203-4108}}%
\email{gfuller@physics.ucsd.edu}
\affiliation{Department of Physics, University of California San Diego, La Jolla, California 92093, USA}

\author{Ian Padilla-Gay\,\orcid{0000-0003-2472-3863}}
\email{ianpaga@berkeley.edu}
\affiliation{Department of Physics, University of California, Berkeley, Berkeley, CA 94720, USA}
\affiliation{Department of Physics, University of California San Diego, La Jolla, California 92093, USA}

\author{Chad T. Kishimoto\,\orcid{0000-0001-7823-0429}}
\email{ckishimoto@sandiego.edu}
\affiliation{Department of Physics and Biophysics, University of San Diego, San Diego, California 92110, USA}

\date{\today}

\begin{abstract}
The collapse of supermassive stars (SMSs, $M\gtrsim10^4\,M_\odot$) to black holes is accompanied by a prodigious flux of neutrinos of all flavors. These are produced thermally via $e^\pm$ annihilations, mostly in the core and just before gravitational trapped surface formation. There, the ratio of fluxes for $\nu_e\bar{\nu}_e$-pairs to $\nu_{\mu}\bar{\nu}_{\mu}/\nu_{\tau}\bar{\nu}_{\tau}$-pairs is $\sim$\,5-to-1. This is because at SMS temperature scales, $\nu_e\bar{\nu}_e$ pairs have both charged and neutral current production channels, whereas $\nu_{\mu}\bar{\nu}_{\mu}/\nu_{\tau}\bar{\nu}_{\tau}$-pairs only have neutral current production channels. We point out that the typical energies of these neutrinos, and the run of density in collapsing radiation-dominated supermassive  configurations, leads to Mikheyev–\allowbreak\hspace{0pt}Smirnov–\allowbreak\hspace{0pt}Wolfenstein (MSW) resonances inside these objects for the atmospheric neutrino mass splitting scale, $\Delta m^2_\mathrm{atm.}\sim2.4\times10^{-3}$ eV$^2$. In the normal neutrino mass hierarchy, adiabatic flavor transformation through the MSW resonances would then swap the fluxes $\nu_e\leftrightharpoons\nu_{\mu,\tau}$, whereas, in the inverted neutrino mass hierarchy, the anti-neutrino fluxes are swapped, $\bar{\nu}_e\leftrightharpoons\bar{\nu}_{\mu,\tau}$. We also examine the prospects for collective neutrino flavor oscillations in these environments. Implications for flavor oscillation's effects on neutrino energy deposition and neutrino-induced nucleosynthesis in the SMS's outer layers are examined, as are prospects for detections of SMS collapses through various means.

\end{abstract}

\maketitle


\section{Introduction}

\lettrine[lines=5]{H}{ere} we point out an intriguing connection between neutrino flavor physics and the conditions and neutrino fluxes accompanying the gravitational collapse of supermassive stars (i.e.,~stars with masses in excess of ${\sim}10^4\,M_\odot$).  The formation and collapse of supermassive stars may figure in the solution of several important cosmological mysteries. Perhaps the most vexing of these is the existence of supermassive black holes (SMBHs) early in the history of the universe~\cite{Mortlock_2011, Venemans_2013, Wu_2015, Ba_ados_2017, jwstsmbh}. Specifically,~\citet{seedobs} suggests that if these objects had existed no later than about 100 million years after the Big Bang and collapsed directly into  black holes at the end of their lives, the high mass ``seeds'' left behind would be able to accrete at the Eddington rate into such massive quasars. Since black holes created from supernovae are at most ${\sim}10\,M_\odot$, they would need to accrete at \emph{super}-Eddington rates for hundreds of millions of years to achieve the same mass as these quasars. This fact could make the scenario involving prompt collapse of individual SMSs and standard Eddington accretion rates preferred. 

Ignoring for now the issue of how such supermassive stars (SMSs) would have formed into single gravitationally bound objects instead of fragmenting, investigating their properties is still of fruitful academic interest at the very least. Moreover, beyond-standard-model (BMS) physics has been invoked as a way for SMSs to circumvent several formation pathway problems. In any case, hydrostatic supermassive objects share some common features~\cite{ds,jwst,lu2024directcollapsesupermassiveblack,PhysRevD.110.083035,freese2025earlyformationsupermassiveblack} regardless of the particulars of their formation mechanisms.

As SMSs are so massive, the vast majority of their pressure support is derived from relativistic particles (in this case, photons) and will therefore eventually suffer general relativistic (GR) instability~\cite{fh1962, iben1963, fowler1964, fww86}. Such configurations can be fully convective and hence have a constant entropy profile throughout the star. In that case, the entropy is determined only by the total mass and overall chemical composition. 

In purely Newtonian gravity, these SMSs would be nearly neutrally buoyant, neither stable nor unstable to radial perturbations. Any contraction or expansion of the star would see the gravitational and pressure forces change together in lock-step and therefore cost no energy, leaving the SMS in a metastable state. Being rather large ($R\sim1$ AU) and diffuse ($\rho_\mathrm{c}\sim10$ g cm$^{-3}$), SMSs' structure can be accurately modeled by just Newtonian theory. 

Their stability (or lack thereof), however, is entirely determined by GR. In general, objects whose pressure is derived primarily from relativistic particles are said to be ``trembling on the verge of instability,''\footnote{Phrase attributed to W.A.~Fowler.} and the inclusion of \nth{1} order GR corrections to Newtonian theory is sufficient to push them over the edge to instability. The inherent non-linearity of GR predicts that both stress-energy \emph{and local spacetime curvature itself} both generate spacetime curvature, which in this context materializes as a slightly stronger than expected gravitational force. Inward radial perturbations would serve to condense the metastable SMS, increasing the pressure forces throughout the star while also deepening the gravitational potential well. While GR at this scale has nothing to say about this increase in pressure, the slightly stronger gravitational forces would then break this delicate balance and shift the configuration from metastable to unstable, leading to a direct collapse into a black hole. This is the so-called Feynman-\allowbreak\hspace{0pt}Chandrasekhar Instability~\cite{fc, Feynman:1996kb}.

The SMS evolutionary timescale from the onset of GR instability to black hole formation is of particular interest to this paper. During collapse, the central density and temperature will increase dramatically until a gravitational trapped surface is formed. Prior to black hole formation, the high densities and temperatures will convert some of the kinetic energy of the infalling stellar material into neutrinos, sapping energy from the interior as they mostly free stream from the core through the SMS basically unimpeded. The chance scatterings on their way out may physically deposit some of their energy into the SMS's outer envelope, or transmute free nucleons into their isospin mirrors thereby also creating $e^-$s and $e^+$s. 

As for neutrino inelastic scattering,~\citet{Fuller_1997} determined that there would be fairly substantial production of deuterium through this mechanism, potentially above the primordial deuterium abundance created during Big Bang Nucleosynthesis (BBN). This could offer tantalizing observational signatures of SMSs having ever existed during the universe's early history.

A possible enhancement of the deuterium production would be realized through the Mikheyev-\allowbreak\hspace{0pt}Smirnov-\allowbreak\hspace{0pt}Wolfenstein (MSW) mechanism, assuming that neutrinos exist in the ``normal mass hierarchy,'' where the lightest mass eigenstate for neutrinos is primarily associated with the $e$ flavor state. Neutrinos already oscillate between the $e$, $\mu$, and $\tau$ flavors in vacuum, however interactions with the electrons in the SMS interior can alter this oscillation. Specifically, $\nu_e$s would ``feel'' both a charged current and neutral current interaction with the ambient electrons that energetics forbid for the other flavor neutrinos, giving the $\nu_e$s a larger in-medium mass that depends on the local electron number density. 

Depending on the neutrino energy, there exists a resonant electron number density at which the flavor oscillation frequency becomes maximal, causing initially $e$-flavor neutrinos produced in the dense core to convert to $\nu_\mu$s and $\nu_\tau$s in the diffuse outer layers. As there is no net positron density, $\bar{\nu}_e$s are unaffected in the normal mass hierarchy. This means that the reaction $\nu_e + n\rightarrow p + e^-$ is now substantially disfavored relative to the reaction $\bar{\nu}_e + p\rightarrow n + e^+$, creating a large population of free neutrons in the SMS. Of course, this only is the case if the MSW resonant density is experienced somewhere within the star, which we will show is the case for a range of SMS progenitor masses. 

The first few sections of this paper will serve as an overview of well established neutrino processes in SMSs, focusing on the formation channels, the cross-sections for interaction on various targets, and the overall structure of SMS. Specifically, Section~\ref{sec:nuCreation} will establish the relative initial amounts of each flavor of neutrino at creation and the total thermally averaged cross section of all neutrinos combined. These considerations will be relevant for determining the interaction rates of each species. 

Section~\ref{sec:bh} will discuss the structure of SMSs pre-instability, the point of instability, and the point at which a gravitational trapped surface is formed. Any analysis in this paper begins at the point of instability and ends when the subsequent black hole appears, as further numerical study is needed to explore conditions that arise after the formation of an event horizon. 

Section~\ref{sec:nfc} will consider all kinds of neutrino flavor oscillation. First, we will consider collective effects that would arise from the high neutrino number densities we expect to occur in SMSs. If collective oscillations are important, they will wash out the initial preference given to $\nu_e\bar{\nu}_e$-pair creation, equalizing all flavors of both neutrinos and antineutrinos. We will show that these non-linear effects are likely unimportant except potentially at the lower end of the SMS mass range. We will then discuss the aforementioned MSW effect, the result of which will suppress the presence of $\nu_e$s relative to $\bar{\nu}_e$s in the SMS's outer layers. We will show that there is an initial mass range of SMSs where the MSW resonances are present within the SMS, and that where it occurs in the SMS also depends on the mass. We will finally show that this flavor oscillation occurs adiabatically, that is, the flavor oscillation wavelength at resonance is much smaller than the resonance width, so the net effect is a true \textit{en mass} flavor conversion.

Section~\ref{sec:cap} will discuss the process of $\bar{\nu}_e$ capture on free protons to convert them to neutrons in the SMS's outer layers. We expect, assuming this transmuted material does not subsequently fall directly into the newly formed black hole, that these free neutrons will readily combine with free protons to form deuterium and potentially even $\alpha$ particles. We then speculate on whether this could provide an observational signature for detection of these SMSs. 

\section{Neutrino Creation and Interactions}\label{sec:nuCreation}

Despite modest temperatures, SMSs posses large entropy-per-baryon ($s\sim100\text{--}1000$), meaning there exists a significant abundance of $e^\pm$-pairs in electromagnetic equilibrium. Neutrinos are created as $\nu\bar{\nu}$-pairs of all flavors in the centers of collapsing SMSs, primarily from relativistic $e^\pm$ annihilations~\cite{fww86, schinder, Fuller_1997}. These can be mediated by the neutral $Z^0$ boson (``neutral current,'' NC) and the charged $W^\pm$ boson (``charged current,'' CC). NC production channels are flavor agnostic, so the $\nu_e\bar{\nu}_e$, $\nu_\mu\bar{\nu}_\mu$, and $\nu_\tau\bar{\nu}_\tau$-pairs are produced in equal amounts. CC production channels only produce $e$ flavor pairs to preserve lepton number of the reactants at tree-level, and the energetics involved forbid $\mu^\pm$ or $\tau^\pm$ annihilation to create the other flavors in the CC channel. This production channel asymmetry leads to a flux enhancement of $\nu_e\bar{\nu}_e$-pairs over $\nu_\mu\bar{\nu}_\mu$ or $\nu_\tau\bar{\nu}_\tau$-pairs, as calculated in Ref.~\cite{shifuller}, to be about 5-to-1, i.e., $\nu_e$-pairs constitute about 70\% of the total neutrino flux. This will be relevant for CC scattering on nucleons, as discussed later in this section.

As neutrinos travel through an SMS, they may scatter on free nuclei, specifically free protons and composite nuclei of atomic mass $A$. Neutrinos may also scatter on electrons, but these are greatly subdominant to nucleon interactions at the energy scales considered in this paper~\cite{fnpa}. NC scatterings are elastic and have the form
\begin{align}
    \stackrel{\scriptscriptstyle(-)}{\nu}_\ell + p &\xrightarrow{Z^0} \stackrel{\scriptscriptstyle(-)}{\nu}_\ell + p,\nonumber\\
    \stackrel{\scriptscriptstyle(-)}{\nu}_\ell + A &\xrightarrow{Z^0} \stackrel{\scriptscriptstyle(-)}{\nu}_\ell + A,\nonumber
\end{align}
where $\stackrel{\scriptscriptstyle(-)}{\nu}_\ell$ refers to neutrinos and anti-neutrinos of lepton flavor $\ell=(e,\mu,\tau)$ and $p$ denotes protons.

CC scatterings are inelastic and have the form
\begin{align}
    \nu_e + n &\xrightarrow{W^\pm} e^- + p,\nonumber\\
    \Bar{\nu}_e + p &\xrightarrow{W^\pm} e^+ + n.\nonumber
\end{align}
Assuming the protons and neutrons are nearly at rest in the COM frame, these CC reactions require $E_\nu>E_\nu^\mathrm{th.}=\Delta m + m_{e^\pm}$ to occur, where $\Delta m\sim\pm1.29\,\mathrm{MeV}$ is the nucleon mass difference after the reaction and $m_{e^\pm}\sim0.511\,\mathrm{MeV}$ is the mass of the resulting electron or positron. The maximum temperatures achieved in SMS collapse (which we will show is ${\sim}0.5\text{--}5\,\mathrm{MeV}$) therefore forbid the other neutrino flavors from participating in the above CC reactions as they do not have the requisite threshold energy to produce the heavy charged leptons.

The NC scattering on nucleons by neutrinos of any lepton flavor has the cross sections~\cite{fnpa,st}
\begin{align}
    \sigma^{\nu p}_\mathrm{NC}(E_\nu) &= \frac{G_\mathrm{F}^2}{4\pi}\left[(1-4\sin^2\theta_\mathrm{W})^2+3g_A^2\right]E_\nu^2, \nonumber \\
    &\approx 2.2\times10^{-44}\pr{\frac{E_\nu}{1\,\mathrm{MeV}}}^2\,\mathrm{cm^2}, \\
    \sigma_\mathrm{NC}^{\nu A}(E_\nu) &= \frac{G_\mathrm{F}^2}{4\pi}A^2 E_\nu^2,\nonumber \\
    &\approx 6.7\times10^{-44}\pr{\frac{A}{4}}^2\pr{\frac{E_\nu}{1\,\mathrm{MeV}}}^2\,\mathrm{cm^2},
\end{align}
where $G_\mathrm{F}\approx1.1664\times10^{-11}\,\mathrm{MeV^{-2}}$ is the Fermi constant (in units of $c=\hbar=1$, \emph{a.k.a.}~``natural'' units), $\sin^2\theta_\mathrm{W}\approx0.23$ is the sine-squared of the Weinberg mixing angle~\cite{ParticleDataGroup:2024cfk}, $g_A\approx1.3$ is the axial-to-vector weak strength ratio for free nucleons, and $E_\nu$ is the neutrino energy. Note that these cross sections are energy dependent, in contrast to, say, the energy independent Thomson cross section ($\sigma_\mathrm{T}\sim6.65\times10^{-25}\,\mathrm{cm^2}$) which governs electromagnetic interactions, and are also ${\sim}$20 orders of magnitude smaller for MeV scale neutrinos.

The CC capture cross section of $\bar{\nu}_e$s on protons is~\cite{fnpa, 1995ApJ...453..792F, 1982ApJ...252..741F}:
\begin{align}
    \sigma_\mathrm{CC}^{\Bar{\nu}_ep} &= \frac{G_\mathrm{F}^2|V_{ud}|^2}{\pi} \pr{1+3g_A^2}(E_\nu-Q)^2, \nonumber\\
    &\approx 10^{-43}\pr{\frac{E_\nu-Q}{1\,\mathrm{MeV}}}^2\,\mathrm{cm^2},
\end{align}
for $E_\nu>Q$, where $|V_{ud}|^2\approx0.948$ is the squared magnitude of the Cabibbo-\allowbreak\hspace{0pt}Kobayashi-\allowbreak\hspace{0pt}Maskawa (CKM) matrix element corresponding to up-down quark flavor mixing~\cite{ParticleDataGroup:2024cfk}, and $Q=m_n-m_p\approx1.29\,\mathrm{MeV}$ is the mass difference between a neutron and a proton. Here, we assume that the electron's momentum after creation is negligible and that there is no phase-space blocking from ambient positrons resulting from Fermi statistics. We will also not consider the cross section for the isospin mirror reaction $\nu_e+n\rightarrow p+e^-$. Any free neutrons would be quickly incorporated into nuclei via strong and electromagnetic interactions. Moreover, the high entropies that attend the collapse of SMSs inhibit the formation of heavy nuclei.

In the context of collapsing SMSs,~\citet{shifuller} found that the energy spectrum of neutrinos produced by annihilating $e^\pm$ pairs near the growing event horizon is well fit by the normalized Fermi-Dirac distribution
\begin{equation}
    f_\nu(E_\nu)= \frac{1}{T_\nu^3 F_2(\eta_\nu)}\frac{E_\nu^2}{e^{E_\nu/T_\nu-\eta_\nu}+1},
\end{equation}
where $T_\nu\approx1.6\,T$ is the ``neutrino temperature,'' about 60\% greater than the local plasma temperature $T$, $\eta_\nu\sim2$ is the ``degeneracy parameter,'' and $F_2(\eta_\nu)$ is a numerical factor calculated in general to be
\begin{equation}\label{eq:dist}
    F_k(\eta_\nu) = \int_0^\infty \frac{x^k}{e^{x-\eta_\nu}+1}\,\dd x,
\end{equation}
hereafter referred to as Fermi integrals of order $k$. It is worth noting that these are fitting parameters and should not be misconstrued as the thermodynamic notions of temperature and degeneracy.

The spectrum-averaged NC elastic scattering cross section is then
\begin{equation}
    \left<\sigma_\mathrm{NC}^{i}\right> = \int_{0}^\infty f_\nu(E_\nu)\sigma_\mathrm{NC}^{i}(E_\nu)\,\dd E_\nu, 
\end{equation}
where $i=\nu p$ or $\nu A$. For SMSs with primordial element abundances, the only other scattering targets besides protons are helium nuclei ($\alpha$ particles) with $A=4$. As a function of the hydrogen mass fraction $X$, the number fractions of each target are
\begin{align*}
    \mathfrak{n}_p(X) &= \frac{4X}{1+3X}, \\
    \mathfrak{n}_\alpha(X) &= 1-\mathfrak{n}_p,
\end{align*}
assuming the SMS is composed of only hydrogen and helium. 

The \emph{total} thermally-averaged NC elastic scattering cross section is then
\begin{align}
    \left<\sigma_\mathrm{NC}\right> &= \mathfrak{n}_p\left<\sigma_\mathrm{NC}^{\nu p}\right> + \mathfrak{n}_\alpha\left<\sigma_\mathrm{NC}^{\nu \alpha}\right>,\nonumber \\
    &= \frac{G_\mathrm{F}^2}{4\pi}\frac{F_4(\eta_\nu)}{F_2(\eta_\nu)}T_\nu^2g(X)
\end{align}
where $g(X) = \mathfrak{n}_p\pr{\left[1-4\sin^2\theta_\mathrm{W}\right]^2+3g_A^2}+16(1-\mathfrak{n}_p)$. For a hydrogen mass fraction of ${\sim}75\%$, typical of a primordial gas, $g(X)\approx5.9$. In standard units, this cross section is therefore
\begin{equation}\label{eq:nccs}
    \left<\sigma_\mathrm{NC}\right> \approx 4.0\times10^{-43} \pr{\frac{T_\nu}{1\,\mathrm{MeV}}}^2\,\mathrm{cm^2}
\end{equation}
for $\eta_\nu=2$. 

To find the thermally averaged cross section for the CC case, we use~\cite{1995ApJ...453..792F}
\begin{equation}
    \left<\sigma_\mathrm{CC}^{\Bar{\nu}_ep}\right> = \int_{E_\nu^\mathrm{th.}}^\infty\sigma_\mathrm{CC}^{\Bar{\nu}_ep}f_\nu(E_\nu)\,\dd E_\nu,
\end{equation}
where $E_\nu^\mathrm{th.}=Q+m_e\sim1.8$ MeV is the threshold energy required for this interaction. This evaluates to
\begin{equation}
    \left<\sigma_\mathrm{CC}^{\Bar{\nu}_ep}\right> \approx 10^{-43}\pr{\frac{T_\nu}{1\,\mathrm{MeV}}}^2G(T_\nu)\,\mathrm{cm^2},
\end{equation}
where the function $G(T_\nu)$ is a linear combination of Fermi integrals of several orders:
\begin{equation}
    G(T_\nu) = \frac{1}{F_2(\eta_\nu)}\sum_{i=0}^4\alpha_i(T_\nu)F_i\pr{\eta_\nu-E_\nu^\mathrm{th.}/T_\nu}
\end{equation}
with temperature dependent coefficients
\begin{align*}
    \alpha_0(T_\nu) &= \pr{Q^2m_e^2+2Qm_e^3+m_e^4}/T_\nu^4,\\
    \alpha_1(T_\nu) &= 2\pr{Q^2m_e+3Qm_e^2+2m_e^3}/T_\nu^3,\\
    \alpha_2(T_\nu) &= \pr{Q^2 + 6Qm_e+6m_e^2}/T_\nu^2,\\
    \alpha_3(T_\nu) &= 2\pr{Q+2m_e}/T_\nu,\\
    \alpha_4(T_\nu) &= 1.
\end{align*}
This function is rather involved but has convenient limiting behaviors. $G(T_\nu)$ vanishes precipitously for $T_\nu\ll Q$, illustrative of the fact that neutrinos with energies sufficient to carry out the reaction are exponentially suppressed. For $T_\nu\gg Q$, $G(T_\nu)$ approaches a limiting value $F_4(\eta_\nu)/F_2(\eta_\nu)$, which for $\eta_\nu=2$ is about 16.

The total thermally-averaged cross section from both the NC and CC interactions is therefore (taking into account that ${\sim}$40\% of neutrinos are $\bar{\nu}_e$):
\begin{align} 
    \left<\sigma\right> &= \left<\sigma_\mathrm{NC}\right> + 0.4\mathfrak{n}_p(X)\left<\sigma_\mathrm{CC}^{\Bar{\nu}_ep}\right>, \nonumber \\
    &\approx 4.0\times10^{-43}\pr{\frac{T_\nu}{1\,\mathrm{MeV}}}^2\left[1+0.1G(T_\nu)\right]\,\mathrm{cm^2},
\end{align}
for $X=0.75$, or in terms of the local plasma temperature $T$:
\begin{equation} \label{eq:sigmatot}
    \left<\sigma\right> \approx 10^{-42}\pr{\frac{T}{1\,\mathrm{MeV}}}^2\left[1+0.1G(1.6T)\right]\,\mathrm{cm^2}.
\end{equation}

This total cross section will determine where, if anywhere, neutrinos are effectively ``trapped'' within the SMS as they stream from the center during collapse. Neutrinos that are not trapped by the time they encounter the MSW flavor resonance will be of particular interest for this paper. 

\section{SMS Structure, Instability, and Black Hole Formation}\label{sec:bh}

The supermassive stars we consider here are radiation-pressure dominated, fully ionized, totally convective, isentropic, and non-rotating hydrostatic configurations of gas with masses in excess of ${\sim}10^4\,M_\odot$. Here, we consider only non-rotating and non-magnetic configurations in order to focus on the neutrino physics of collapse. For these SMSs, the ratio of gas pressure to total pressure $\beta$ is constant throughout the star, and is calculated to be~\cite{beta}
\begin{equation}
    \beta\equiv\frac{P_\mathrm{gas}}{P_\mathrm{rad.}+P_\mathrm{gas}}\approx\frac{4.3}{\mu}\pr{\frac{M}{M_\odot}}^{-1/2},
\end{equation}
where $\mu\approx0.59$ is the mean molecular weight for a fully ionized primordial gas, and $P_\mathrm{gas}$ and $P_\mathrm{rad.}$ are the Maxwell-Boltzmann pressure and radiation pressure contributions, respectively. $\beta$ is much less than unity for the mass range considered in this paper ($10^4\lesssim M/M_\odot\lesssim10^8$). The entropy-per-baryon $s$ is similarly calculated as
\begin{equation}
    s\approx\frac{4}{4.3}\pr{\frac{M}{M_\odot}}^{1/2} - \frac{4}{\mu},
\end{equation}
in units of Boltzmann's Constant $k_\mathrm{B}$. 

The above properties combine to make an SMS's equation of state match closely to that of a polytrope, i.e.:
\begin{equation}\label{eq:eos}
    P(s, \rho) = K(s)\rho^{1+1/n},
\end{equation}
where $P$ is the pressure, $\rho$ is the mass density, $n$ is the ``polytropic index,'' and $K(s)$ is the ``polytropic constant'' that depends on the entropy-per-baryon. Since $s$ is constant throughout the SMS, $K(s)$ is also a constant. Specifically, we take $n\sim3$ for SMSs, so $K(s)$ is calculated to be
\begin{align}
    K(s) &= \frac{1}{1-\beta}\frac{a}{3}\pr{\frac{3}{4a}\frac{s}{ m_\text{b}}}^{4/3}, \nonumber \\ 
    &\approx\frac{0.2843}{1-\beta}\pr{\frac{s}{1000}}^{4/3}\,\rm{MeV}^{-4/3},
\end{align}
where $a=\pi^2/15$ is the radiation constant and $m_\text{b}\approx938$~MeV is the mass of a baryon.

The total energy for such a configuration is given by~\cite{mnras}
\begin{multline}\label{eq:totalE}
    E(\rho_\mathrm{c}) = k_1MK(s)\rho_\mathrm{c}^{1/n} - k_2GM^{5/3}\rho_\mathrm{c}^{1/3}\\ - k_4G^2M^{7/3}\rho_\mathrm{c}^{2/3},
\end{multline}
where $k_1\approx1.7558$, $k_2\approx0.6390$, and $k_4\approx0.9183$ are numerical constants calculated from integrating over an $n=3$ polytrope, $\rho_\mathrm{c}$ is the central density, and $G\sim6.7\times10^{-45}\,\mathrm{MeV^{-2}}$ is Newton's gravitational constant. 

The terms represent, in order, the internal kinetic energy (non-relativistic Maxwell-Boltzmann and relativistic photon contributions), the gravitational binding energy, and first order corrections from GR, all as a function of $M$ and $\rho_\mathrm{c}$. For $n\sim3$, the virial theorem dictates that the first two terms nearly perfectly cancel, so the total energy is dominated by the GR correction term. Since that energy is proportional to $G^2$, it is in effect a gravitational ``self-coupling,'' hinting at the true non-linear nature of GR. 

As these SMSs are radiation dominated, they will also radiate photons from their surface at the Eddington limited rate $L_\mathrm{Edd.}$~\cite{st}, defined to be
\begin{align}
    L_\mathrm{Edd.} &= \frac{4\pi GMm_\mathrm{b}}{\sigma_\mathrm{T}}\nonumber,\\
    &\approx1.3\times10^{38}\pr{\frac{M}{M_\odot}}\,\mathrm{erg\,s^{-1}}.
\end{align}
This energy loss causes the SMS to quasi-statically contract, raising the central density and temperature over time. 

\subsection{SMS Structure and Instability}\label{ss:poly}

With the above equation of state (Eq.~\eqref{eq:eos}, $n=3$) and the following rescalings for radius $r$ and density:
\begin{align}
    r &= \alpha_\mathrm{LE} \xi = \pr{\frac{K}{\pi G}}^{1/2}\rho_\mathrm{c}^{-1/3}\xi, \\
    \rho &= \rho_\mathrm{c}\varphi^3,
\end{align}
the equations for hydrostatic equilibrium can be manipulated to yield the Lane-Emden Equation of order 3:
\begin{subequations}
    \begin{align}\label{eq:le}
        \frac{1}{\xi^2}\frac{\dd}{\dd\xi}\pr{\xi^2\frac{\dd\varphi}{\dd\xi}} + \varphi^3 &= 0, \\
        \varphi(0) &= 1, \\
        \varphi^\prime(0) &= 0,
    \end{align}
\end{subequations}
the solution of which is used to find the density profile as a function of radius. Denoting the location of the first zero of $\varphi(\xi)$ as $\varphi(\xi_3)=0$, the total mass $M$ and radius $R$ of the configuration are
\begin{align}
    M &= 4\pi\pr{\frac{K}{\pi G}}^{3/2}\xi_3^2\left|\varphi^\prime(\xi_3)\right|\label{eq:mass}, \\ 
    R &= \pr{\frac{K}{\pi G}}^{1/2}\rho_\mathrm{c}^{-1/3}\xi_3, \label{eq:Rle}
\end{align}
where $\xi_3\approx6.9868$ and $|\varphi^\prime(\xi_3)|\approx0.0424$.

Under a purely Newtonian theory of gravity, these $n=3$ polytropes are neutrally stable, i.e.~it costs no energy to expand or contract the configuration. This fact is also realized by the hydrostatic mass in Eq.~\ref{eq:mass} having no dependence on the central density, lending them a self-similar structure.

Upon considering first order GR corrections to Newtonian theory, one can derive the critical central density $\rho_\mathrm{c,\,crit.}$ above which the configuration is unstable to collapse~\cite{beta,st}:
\begin{equation}\label{eq:rhoccrit}
    \rho_\mathrm{c,\,crit.}\approx3.98\pr{\frac{\mu}{0.59}}^{-3}\pr{\frac{M}{10^5\,M_\odot}}^{-7/2}\,\mathrm{g\,cm^{-3}}.
\end{equation}

Per~\citet{st}, the SMS will contract from a diffuse state ($\rho_\mathrm{c}\sim0$) to the critical state on roughly the Kelvin-Helmholtz thermal timescale $\tau_\mathrm{KH}$. For $n\sim3$ polytropes radiating at the Eddington luminosity $L_\mathrm{Edd.}$, this timescale is calculated using the final term in Eq.~\eqref{eq:totalE} at the critical state via
\begin{equation}
    \tau_\mathrm{KH}\sim\frac{k_4G^2M^{7/3}\rho_\mathrm{c,\,crit.}^{2/3}}{L_\mathrm{Edd.}}\approx8.9\times10^3\pr{\frac{M}{10^5\,M_\odot}}^{-1}\,\mathrm{yr},
\end{equation}
since the internal energy and Newtonian gravitational binding energy cancel each other out.

During this quasi-static contraction phase, the central density will approach $\rho_\mathrm{c,\,crit.}$ from below (Eq.~\eqref{eq:rhoccrit}), at which point it will become unstable to collapse and begin a rapid contraction phase roughly at the free-fall timescale $\tau_\mathrm{ff}\sim(G\rho)^{-1/2}\approx4.8\times10^{-5}(M/10^5\,M_\odot)^{7/4}\,\mathrm{yr}$. For $M\sim10^8\,M_\odot$, both of the prior timescales are equal, so masses greater than this are not considered as there is basically no distinction between a quasi-static contraction phase and free-fall collapse.

\subsection{Trapped Surface Formation and Neutrino Trapping}\label{sec:trap}

As the SMS collapses after instability sets in, neutrino production in the center will sap entropy from the SMS. Ref.~\cite{fww86} shows that this loss in entropy will cause a smaller so-called ``homologous core'' (HC) with mass $M^\mathrm{HC}\sim0.1M^\mathrm{init.}$ to collapse while maintaining the same overall $n=3$ polytropic structure. The central density will rise and the radius will decrease per Eq.~\eqref{eq:Rle}, while the total HC mass stays the same. This process continues until a trapped surface is formed at a radius $r_\bullet$ such that
\begin{equation}
    r_\bullet = 2GM_\mathrm{enc.}(r_\bullet)
\end{equation}
where $M_\mathrm{enc.}(r)$ is the mass enclosed at radius $r$ with boundary conditions $M_\mathrm{enc.}(0)=0$ and $M_\mathrm{enc.}(R) = M$. 

The enclosed mass function can be calculated directly to be, using the Lane-Emden rescalings:
\begin{align}
    M_\mathrm{enc.}(r) &= \int_0^r4\pi (r^\prime)^2 \rho(r^\prime)\,\dd r^\prime,\nonumber\\    
     &=4\pi\alpha_\mathrm{LE}^3\rho_\mathrm{c}\xi^2\left|\varphi^\prime(\xi)\right|.
\end{align}
The condition for a trapped surface now simplifies to the transcendental equation
\begin{equation}\label{eq:bhole}
    \xi_\bullet = 8 K(s)\rho_\mathrm{c}^{1/3}\xi_\bullet^2 \left|\varphi^\prime(\xi_\bullet)\right|,
\end{equation}
where for fixed entropy-per-baryon $s$, the above equation is just parameterized by $\rho_\mathrm{c}$.

There exists some minimum central density $\rho_\bullet$ such that Eq.~\eqref{eq:bhole} has one solution for $\xi_\bullet$ for any choice of HC mass. It turns out that, for an $n=3$ polytrope, $\xi_\bullet$ is independent of the mass and always occurs at $\xi_\bullet\approx2.632$, a bit under 40\% of the total radius. No GR considerations went into this calculation, so it should be emphasized that this ``quick-and-dirty'' approach is likely an over-estimate of both the central density at trapped surface formation and the subsequent trapped surface radius. 

Indeed,~\citet{linke} have already performed several full-GR simulations of collapses of intially $n=3$ polytropes with masses ranging from $5\times10^5\,M_\odot$ to $10^9\,M_\odot$. They found that---regardless of the mass---a gravitational trapped surface forms during collapse that encloses the innermost ${\sim}25\%$ of the total SMS mass. Assuming the collapsing SMS structure is still close to that of an $n=3$ polytrope (this is surely not the case when GR considerations become important), this means the trapped surface forms at about $20\%$ of the total radius, which would make our na\"ive calculation an overestimate for the trapped surface size. The authors, however, do not report the central densities at which these trapped surfaces form, so for the purpose of this paper the $\rho_\bullet$ determined in Eq.~\eqref{eq:bhole} will be sufficient. The densities that satisfy that condition for various SMS homologous core masses have been computed numerically and are presented in Table~\ref{tab:den}.

Despite the small cross-section for NC elastic scattering and CC inelastic scattering (Eq.~\eqref{eq:sigmatot}), the high densities and temperatures involved behoove us to consider if the neutrinos become ``trapped'' within the SMS. We will consider the neutrinos trapped if their mean free path $\lambda_\mathrm{mfp}$ is of order the radius $R$ and determine at what $\rho_\mathrm{c}$ this is the case. This condition yields the equation:
\begin{equation}
    \lambda_\mathrm{mfp} \sim R(\rho_\mathrm{\nu\,trap}) \sim \frac{\mu m_\mathrm{b}}{\rho_\mathrm{\nu\,trap} \left<\sigma(\rho_\mathrm{\nu\,trap})\right>},
\end{equation}
where $R(\rho_\mathrm{c})$ is given in Eq.~\eqref{eq:Rle}.

The cross section's dependency on central density is determined from noting that a radiation dominated $n=3$ polytrope's central temperature is related to the central density via 
\begin{equation}\label{eq:Tc}
    T_\mathrm{c}=\pr{\frac{3K}{a}\pr{1-\beta}}^{1/4}\rho_\mathrm{c}^{1/3},
\end{equation}
where $T_\nu \approx 1.6\,T_\mathrm{c}$ as discussed previously. The neutrino trapping densities $\rho_\mathrm{\nu\,trap}$ are also calculated numerically and are presented in Table~\ref{tab:den}.

\begin{table}
\caption{\label{tab:den} Calculated central densities at which a trapped surface is formed ($\rho_\bullet$) and at which the neutrinos' mean free path becomes less than the radius ($\rho_\mathrm{\nu\,trap}$). For homologous core masses ${\gtrsim}10^5\,M_\odot$, the black hole formation density is lower than the neutrino trapping density, meaning that neutrinos are free streaming from the core to the surface before a trapped surface is formed.}
\begin{tabular}{@{}ccc@{}}
\toprule
$M^\mathrm{HC}$ [$M_\odot$] & $\rho_\bullet$ [g/cm$^3$] & $\rho_\mathrm{\nu\,trap}$ [g/cm$^3$] \\ \midrule
$10^4$           & $1.61\times10^9$          & $4.89\times10^7$                \\
$10^5$           & $1.52\times10^7$          & $1.52\times10^7$                \\
$10^6$           & $1.50\times10^5$          & $4.81\times10^6$                \\
$10^7$           & $1.49\times10^3$          & $1.52\times10^6$                \\
$10^8$           & $1.49\times10^1$          & $4.80\times10^5$                \\ \bottomrule
\end{tabular}
\end{table}

It is clear that for some homologous core mass ${\sim}10^5\,M_\odot$, the neutrinos are trapped by scatterings at the same time as the formation of a gravitational trapped surface. Therefore, any SMS mass above this will permit neutrinos to escape the star even after the trapped surface is formed.

\section{Neutrino flavor conversion}\label{sec:nfc}

Neutrino flavor conversion is a complex quantum mechanical phenomenon that can have important consequences on the macroscopic scale. Dense astrophysical environments---such as core-collapse supernovae and neutron stars mergers---are two important venues where neutrinos are expected to undergo flavor conversion because of extremely high neutrino number densities near the neutrino-baryon decoupling regions. Neutrino forward scattering on background matter offers a simple phenomenology described by the MSW effect; however, neutrino forward scattering on background neutrinos can lead to a plethora of non-linear collective effects where neutrinos of different momenta couple to each other significantly~\cite{Duan:2010bg}. 
These non-linear interactions lend a great deal of complexity to the oscillation phenomenology. Here we explore whether neutrino collective effects are important for SMSs.

In order to describe the oscillation phenomenology, one needs to know the energy scales of the Hamiltonian in the quantum kinetic equations (QKEs), which reads
\begin{equation}
    H = H_\mathrm{vac.}+H_\mathrm{matt.}+H_{\nu\nu}.
\end{equation}
The vacuum term is proportional to the vacuum oscillation frequency
\begin{equation}
    \omega = \frac{\Delta m^2}{2E_{\nu}}\sim6.3\pr{\frac{1\,\mathrm{MeV}}{E_\nu}}\,\mathrm{km}^{-1},
\end{equation}
where $\Delta m^2=m_2^2-m_1^2$ is the difference between the squared mass eigenvalues in vacuum, $E_\nu$ is the neutrino energy, and we used $\Delta m^2\sim10^{-3}\,\mathrm{eV^2}$. This is roughly the scale for the wavenumber of flavor oscillations in vacuum. For the following section of this paper, wherein we discuss the matter term in the Hamiltonian in more detail, we will take $\Delta m^2=\Delta m^2_\mathrm{atm.}\approx2.4\times10^{-3}\,\mathrm{eV}^2$, the so-called ``atmospheric'' mass-squared splitting. 

The matter term is only felt by $\nu_e$s, and it interferes with the vacuum contribution to produce the MSW effect. It is proportional to the neutrino-electron interaction strength, with its scale being set by
\begin{equation}
    \lambda=\sqrt{2}G_{F}n_{e},
\end{equation}
where $n_{e}=\rho/\mu m_\mathrm{b}$ is the local number density of electrons. In the dense core, just before gravitational trapped surface formation, we expect $\lambda\gg\omega$, whereas at the relatively diffuse SMS outer layers we expect $\lambda\ll\omega$. Somewhere within the collapsing SMS homologous core, we expect $\lambda\sim\omega$, wherein the interesting MSW effect lies. A detailed exploration of this effect is provided in the forthcoming second, third, and fourth sub-sections of this section. 

Before discussing the MSW effect, we will discuss collective flavor oscillations borne from the $H_{\nu\nu}$ term in the Hamiltonian and whether we expect these non-linear interactions to be important in collapsing SMS homologous cores.

\subsection{Collective Flavor Oscillations}

The Hamiltonian contribution that describes neutrino forward scattering on other neutrinos is fixed by the neutrino-neutrino interaction strength $\mu_{\nu\nu}=\sqrt{2}G_F n_{\nu}$, where $n_{\nu}$ is the number density of neutrinos. For a collapsing homologous core,~\citet{shifuller} calculated the peak luminosity of $e^+e^-$ pair-produced neutrinos to be $L_\nu\sim10^{56}\pr{M^\mathrm{HC}/10^5\,M_\odot}^{-3/2}\,\mathrm{erg\,s^{-1}}$. 

Take for example a collapsing $10^5\,M_\odot$ homologous core that forms a trapped surface at $R_\bullet\sim2GM^\mathrm{HC}$. The neutrinos have average energy $\left<E_\nu\right>\sim1\,\mathrm{MeV}$ (since $E_\nu\sim T_\nu\sim T_\mathrm{c}$ calculated from using $\rho_\mathrm{c}=\rho_\bullet$ for $M^\mathrm{HC}=10^5\,M_\odot$ in Eq.~\eqref{eq:Tc}), so the neutrino number density produced at the gravitational trapped surface $R_\bullet\sim10^{11}\,\mathrm{cm}$ is $n_\nu\sim L_\nu/(4\pi cR_\bullet^2 E_\nu)\sim10^{28}\,\mathrm{cm^{-3}}$. We can now calculate the neutrino-neutrino interaction strength scale $\mu_{\nu\nu}$ to be
\begin{equation}
    \mu_{\nu\nu} \simeq \frac{\sqrt{2}G_F}{4\pi c R_\bullet^2}\frac{L_\nu}{E_\nu} \sim 10\,{\rm km^{-1}}.
\end{equation}
Thus, one can see that roughly $\mu_{\nu\nu}\sim \omega$, meaning that collective effects, if any, would occur on the same timescale as neutrino vacuum oscillations. SMS masses larger than $10^5\,M_\odot$ see a precipitous drop in $\mu_{\nu\nu}$ relative to $\omega$ (both $L_\nu$ and $1/R_\bullet^2$ decrease quickly with increasing mass), meaning that collective effects are even less prominent.

Moreover, the neutrino emission derived entirely from annihilating thermal $e^+e^-$ pairs, so the spatial distribution of these neutrinos is completely isotropic. Therefore, there are no crossings in the (anti)neutrino angular distributions, suggesting that fast flavor conversion (FFC) mechanisms~\cite{Sawyer2005a,Sawyer2009a,Sawyer2016a} (see~\cite{Tamborra:2020cul, Johns:2025mlm} for recent reviews) are most likely not operative~\cite{morinaga2022fast,Dasgupta2022a,Dasgupta:2025quc}. Slow flavor conversion (SFC), on the other hand, can lead to unstable solutions of the flavor evolution even for isotropic systems~\cite{Duan:2010bg, Hannestad:2006nj} and non-isotropic systems without angular crossings~\cite{Padilla-Gay:2025tko}. If the distributions are perfectly isotropic, the system is described by the ``single-angle approximation'' which is unstable in the inverted mass ordering, i.e.~$\omega\propto\Delta m^2 < 0$, and stable in the normal mass hierarchy $\omega >0$~\cite{Duan:2010bg, Hannestad:2006nj}. In reality, however, the neutrino distributions are not exactly isotropic, and even small perturbations in the emission can lead to multi-angle instabilities that break the axial symmetry~\cite{Raffelt:2013rqa}. The single-energy Hamiltonian describing neutrino-neutrino forward scattering is
\begin{equation}
    H_{\nu\nu} = \mu_{\nu\nu} \int dv^{\prime} \pr{\hat{\rho}^{\prime}-\alpha \hat{\bar{\rho}}^{\prime}}(1-v v^{\prime}),
\end{equation}
where $\alpha=n_{\bar{\nu}_e}/n_{\nu_e}$ is the asymmetry parameter ($\alpha=1$ in SMSs) and $v=\cos{\theta_\nu}$ is the cosine of the neutrino emission angle. $\hat{\rho}$ and $\hat{\bar{\rho}}$ are the neutrino and anti-neutrino density matrices, respectively, whose diagonal terms are proportional to the neutrino number densities and the off-diagonal terms encode coherence among flavors.

Lastly, a novel type of instability, the so-called collisional flavor instability (CFI), has been identified in core-collapse supernovae and neutron star mergers~\cite{Johns2021_1,Padilla-Gay:2022wck,Lin:2022dek,Xiong:2022zqz,Liu:2023pjw}. This type of instability, arises due to sufficiently distinct scattering rates on matter of $\nu_e$ and $\bar{\nu}_e$ and is most effective when $n_{\nu_e}\approx n_{\bar{\nu}_e}$, both conditions present in SMSs. SMSs can guarantee a large asymmetry between collision rates $\Gamma_{\nu_e}\sim n\sigma_{\rm NC}$ and $\Gamma_{\bar{\nu}_e}\sim n(\sigma_{\rm NC}+\sigma_{CC})$, where $n$ is the number density of targets (roughly the number density of protons). These can differ by as much as $\Gamma_{\bar{\nu}_e}/\Gamma_{{\nu}_e}\simeq 4$ despite the exponential temperature cutoff dependence for $T\lesssim 1.8\,\mathrm{MeV}$ in $\sigma_\mathrm{CC}$. 

The collision rates for neutrinos in SMSs are, however, very small, reaching only $\Gamma_{\nu_e},\Gamma_{\bar{\nu}_e}\sim10^{-3} \ {\rm km^{-1}}$ in the dense core for even the most favorable low mass scenario with $M^{\rm HC}\sim10^{4}\,M_{\odot}$. Higher mass SMS homologous cores are less dense at gravitational trapped surface formation, and therefore less hot, lowering these scattering rates and their asymmetry. In SMSs, we can therefore conclude that collisional rates are so low that CFI is most likely unimportant even where it does arise. We leave a more thorough exploration of flavor instabilities in lower mass SMSs for future work. 

All of that said, in this paper we can rather safely say that collective effects are not important. We will assume the normal mass hierarchy, so since $\mu_{\nu\nu}\sim \omega$ and because our neutrino distributions are perfectly isotropic (single-angle), no flavor instabilities will occur. The scattering rates on matter for both $\nu_e$s and $\bar{\nu}_e$s are small, so collisional flavor instabilities are likely unimportant. These considerations indicate that there is basically no influence from collective effects in general, so in-medium flavor conversion will proceed exclusively through the MSW effect.

\subsection{MSW Flavor Conversion}\label{sec:msw}

The primary neutrino production channel in collapsing SMSs is $e^+ + e^-\rightarrow\nu+\bar{\nu}$. This produces neutrino/anti-neutrino pairs of any flavor. If the reaction is mediated by a $Z^0$ boson, then the flavors are produced in equal amounts. If the mediator is a $W^\pm$ boson, however, $\nu_e\bar{\nu}_e$-pairs will be produced to preserve lepton number at tree-level. As discussed previously, this causes a flavor asymmetry of about 5-to-1 in the relative numbers of $\nu_e\bar{\nu}_e$-pairs over $\nu_\mu\bar{\nu}_\mu$ and $\nu_\tau\bar{\nu}_\tau$-pairs.

After being created, neutrinos will oscillate between the $e$, $\mu$, and $\tau$ flavors as they propagate through matter and vacuum. The former case is of particular interest to this paper as there exists a matter density---depending on the neutrino energy---such that this flavor mixing is maximal, \emph{a.k.a.}~the MSW effect. For simplicity, we will consider a scenario where there are only two neutrino flavors (weak eigenstates) $\ket{\nu_e}$ and $\ket{\nu_x}$, the latter representing both the $\mu$ and $\tau$ flavors. Neutrinos also have mass eigenstates $\ket{m_1}$ and $\ket{m_2}$ where $m_1<m_2$. In this 2$\times$2 case, we take the unitary transformation between flavor states and mass states in vacuum to be:
\begin{subequations}
    \begin{align}
        \ket{\nu_e} &= \cos\theta\ket{m_1} + \sin\theta\ket{m_2},\\
        \ket{\nu_x} &= -\sin\theta\ket{m_1} + \cos\theta\ket{m_2},
    \end{align}
\end{subequations}
where $\theta\neq0$ is the vacuum mixing angle. Here, we will take $\theta=\theta_{13}\approx0.15$~\cite{ParticleDataGroup:2024cfk} so that the electron flavor is primarily in the smaller mass eigenstate in vacuum, i.e.~the normal mass hierarchy. Neutrinos are always created in flavor eigenstates to preserve lepton number, but the Schr\"odinger equation predicts that $\ket{m_1}$ and $\ket{m_2}$ will accumulate different complex phases as they propagate since $m_1\neq m_2$.

The situation becomes more complicated in the presence of matter. While both flavors can forward scatter on electrons via NC $Z^0$ exchange:
\begin{align*}
    \nu_e + e^- &\xrightarrow{Z^0} \nu_e + e^-,\\
    \nu_x + e^- &\xrightarrow{Z^0} \nu_x + e^-,
\end{align*}
the lower mass of the electron also allows charged current $W^\pm$-mediated forward scattering between electrons and $\nu_e$ only:
\begin{equation*}
    \nu_e + e^- \xrightarrow{W^\pm} e^- + \nu_e,
\end{equation*}
in effect giving the $\nu_e$s an extra charged current potential $A_\mathrm{CC}$. One can derive this extra potential to be~\cite{fnpa}
\begin{equation}
    A_\mathrm{CC}=2\sqrt{2}G_\mathrm{F}E_\nu n_e,
\end{equation}
where $n_e$ is the electron number density. This additional potential is completely unfelt by the $\nu_x$s as they propagate.

Similar to how photons traveling in a medium acquire an effective mass from electromagnetic interactions with the ambient electrons, the $\nu_e$s now have an effective mass that increases with increasing $A_\mathrm{CC}$. A higher electron number density forces the $\nu_e$s to be mostly in the heavier mass eigenstate $\ket{m_2}$ instead of the lighter mass eigenstate $\ket{m_1}$, meaning that the in-medium mixing angle $\theta_\mathrm{M}$ now depends on the surrounding matter. This matter dependent mixing manifests as the superposition
\begin{subequations}
    \begin{align}
        \ket{\nu_e} &= \cos\theta_\mathrm{M}\ket{\nu_1} + \sin\theta_\mathrm{M}\ket{\nu_2},\label{eq:nuem}\\
        \ket{\nu_x} &= -\sin\theta_\mathrm{M}\ket{\nu_1} + \cos\theta_\mathrm{M}\ket{\nu_2}\label{eq:nuxm},
    \end{align}
\end{subequations}
where $\theta_\mathrm{M}$ is related to the vacuum mixing angle $\theta$ and the CC potential $A_\mathrm{CC}$ via~\cite{fnpa}
\begin{equation}\label{eq:cos2thetaM}
    \cos2\theta_\mathrm{M} = \frac{\Delta m^2\cos2\theta - A_\mathrm{CC}}{\sqrt{\pr{\Delta m^2\cos2\theta-A_\mathrm{CC}}^2+\pr{\Delta m^2\sin2\theta}^2}}
\end{equation}
where $\Delta m^2=m_2^2-m_1^2$ is the difference of the squares of the mass eigenvalues in vacuum. As stated previously, we will take $\Delta m^2=\Delta m^2_\mathrm{atm.}\approx2.4\times10^{-3}\,\mathrm{eV^2}$.

\subsection{MSW Resonance Density}

For dense environments where $A_\mathrm{CC}\gg\Delta m^2\cos2\theta$, Eq.~\eqref{eq:cos2thetaM} suggests that $\cos2\theta_\mathrm{M}\rightarrow-1$, hence $\theta_\mathrm{M}\rightarrow\pi/2$. Therefore,  the superposition in Eqs.~\eqref{eq:nuem} and \eqref{eq:nuxm} is such that $\ket{\nu_e}$ is primarily in the heavy mass state and $\ket{\nu_x}$ is primarily in the lighter state. Outside of the SMS, where $n_e\sim0$ so $A_\mathrm{CC}\ll\Delta m^2\cos2\theta$, it is instead the case that $\cos2\theta_\mathrm{M}\rightarrow\cos2\theta$, so the superposition reduces to the standard vacuum case. 

Between these two extremes, where $A_\mathrm{CC}\sim\Delta m^2\cos2\theta$, the numerator vanishes so $\cos2\theta_\mathrm{M}\sim0$, meaning that $\theta_\mathrm{M}\sim\pi/4$. This condition corresponds to maximal mixing between the two flavor states which establishes a ``resonant density'' $n_e^\mathrm{res.}$ given by 
\begin{align}\label{eq:res}
    n_e^\mathrm{res.} &= \frac{\Delta m^2 \cos2\theta}{2\sqrt{2}G_\mathrm{F}E_\nu},\nonumber\\
    &\approx 9.6\times10^{27}\cos2\theta\pr{\frac{E_\nu}{1\,\mathrm{MeV}}}^{-1}\,\mathrm{cm^{-3}},
\end{align}
which corresponds to a mass density of
\begin{align}
    \rho_\mathrm{res.} &= \frac{2}{1+X}m_\mathrm{b}n_e^\mathrm{res.}, \nonumber \\
    &\approx 1.9\times10^4\cos2\theta\pr{\frac{E_\nu}{1\,\mathrm{MeV}}}^{-1}\,\mathrm{\frac{g}{cm^3}},
\end{align}
where $X\approx0.75$ is the mass fraction of hydrogen in a primordial gas cloud. 

Averaging over the distribution described in Eq.~\eqref{eq:dist} we find that the resonant density in terms of the plasma temperature $T$ is
\begin{equation}
    \left<\rho_\mathrm{res.}\right> \approx4.1\times10^3 \cos2\theta \pr{\frac{T}{1\,\mathrm{MeV}}}^{-1}\,\frac{\mathrm{g}}{\mathrm{cm}^3},
\end{equation}
where we have again used $T_\nu\approx1.6\,T$ as per Ref.~\cite{shifuller}. 

\begin{figure}
    \centering
    \includegraphics[width=\linewidth]{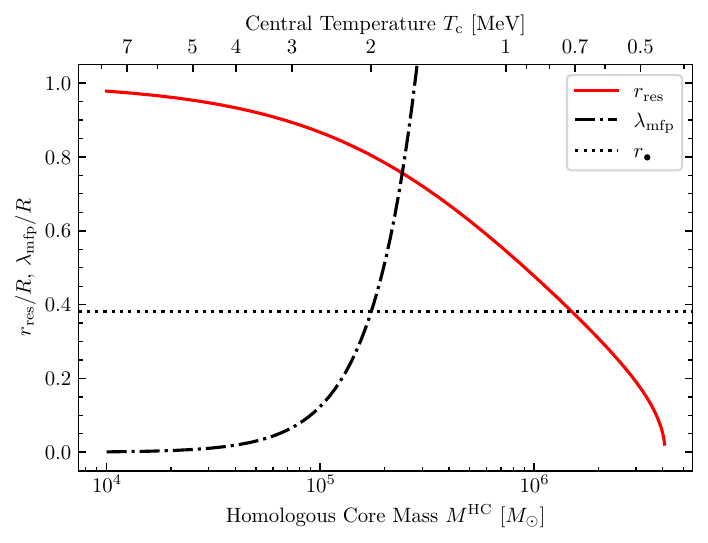}
    \caption{The radial location of the resonant density $\left<\rho_\mathrm{res.}\right>$ (solid red line, scaled by the HC radius) at the black hole forming central density $\rho_\bullet$ as a function of the HC mass $M^\mathrm{HC}$ on the bottom $x$-axis and the central temperature $T_\mathrm{c}$ on the top $x$-axis. The mean free path of neutrinos produced at the center as a ratio of the HC radius is also plotted as the dashed-dotted black line. We have chosen the neutrino mass-splitting and mixing angle to be, respectively, $\Delta m^2=\Delta m^2_\mathrm{atm.}\approx2.4\times10^{-3}$ eV$^2$ and $\theta=\theta_{13}\approx0.15$. The radial location of the trapped surface (per Eq.~\eqref{eq:bhole}) as a fraction of the HC radius is plotted as a dotted black line.}
    \label{fig:rres}
\end{figure}

A plot of where this resonance occurs in the homologous core at the black hole forming density $\rho_\bullet$ as a function of homologous core mass is in Figure~\ref{fig:rres}. The plasma temperature $T$ was taken to be the central temperature $T_\mathrm{c}$, related to $\rho_\bullet$ by Eq.~\eqref{eq:Tc}. Also plotted in the same figure is the neutrino mean free path at $\rho_\bullet$ for neutrinos emitted from the core. The region following the intersection of these two curves is of interest as those neutrinos would undergo MSW resonance and are likely to escape the homologous core without being trapped by scatterings. A final addition to the figure is the location of the gravitational trapped surface at $\rho_\bullet$ using the na\"ive prescription detailed in Section~\ref{sec:trap}. While the precise location of this curve should be calculated with an approach that tackles the collapse with full GR, it should still be a guide to the eye for where effects from the MSW resonances are suppressed as material may be trapped behind an event horizon.

\subsection{Adiabatic Flavor Conversion}

Analogous to the geometric optics limit for light waves and lenses/mirrors, \emph{en masse} flavor conversion between neutrino species in non-uniform media is only observed if the length scale on which the density changes is much larger than the flavor oscillation wavelength. The location where this matters most is at densities around $\rho_\mathrm{res}$ since the flavor mixing is maximal and the in-media mass-squared difference, $\Delta m^2_\mathrm{M}$, is minimal:
\begin{align}
    \Delta m^2_\mathrm{M} &= \sqrt{(\Delta m^2\cos2\theta - A_\mathrm{CC})^2+(\Delta m^2\sin2\theta)^2} ,\nonumber \\
     &\stackrel{\text{res.}}{=} \Delta m^2 \sin2\theta.
\end{align}

Following the prescription in~\citet{qianfuller1995}, we can introduce an ``adiabaticity parameter'' at the resonant density, notated as $\gamma_\mathrm{res.}$, given by
\begin{equation}
    \gamma_\mathrm{res}=\frac{\Delta m^2}{2E_\nu}\frac{\sin^2 2\theta}{\cos2\theta}\left|\frac{1}{\rho}\frac{\dd \rho}{\dd r}\right|^{-1}_\mathrm{res.},
\end{equation}
which is the ratio of the resonance width $\delta r=|\mathrm{d}\ln(\rho)/\mathrm{d}r|^{-1}\tan2\theta\sim R\tan2\theta$ to the flavor oscillation wavelength at the resonance density. The probability that a neutrino will ``jump'' to the other mass state at resonance is $P\sim\exp\pr{-\tfrac{\pi}{2}\gamma_\mathrm{res.}}$~\cite{haxton1987}. If the neutrino does indeed jump from one mass state to the other at the resonant density, there will be no flavor conversion, i.e.~the correlation between the flavor and mass states remains the same between the high density and vacuum environments.

On the other hand, if $\gamma_\mathrm{res}\gg1$, the probability of a mass state jump is negligible and \emph{en masse} flavor transformation will occur. If we calculate the adiabaticity parameter for $M^\mathrm{HC}=10^4\,M_\odot$, we find $\gamma_\mathrm{res}\sim500$, so the jump probability is exponentially suppressed. For masses greater than $10^4\,M_\odot$, the neutrino energy decreases and the density scale height increases, so $\gamma_\mathrm{res}$ increases in turn and the jump probability continues to be exponentially suppressed.

If a $\nu_e$ is created in an initially dense environment where $\cos2\theta_\mathrm{M}\sim-1$, passes through the resonance density in Eq.~\eqref{eq:res}, then continues out to where $\cos2\theta_\mathrm{M}\sim\cos2\theta$, the probability that it remains in the $\nu_e$ state upon interacting with another particle is found to be approximately~\cite{fnpa}
\begin{equation}\label{eq:survival}
    P_{\nu_e\rightarrow\nu_e} \approx \frac{1}{2}\pr{1 + \cos2\theta_{\mathrm{M}}^{\rm (i)}\cos2\theta},
\end{equation}
if $\gamma_\mathrm{res.}\gg1$, where $\theta_\mathrm{M}^\mathrm{(i)}$ is the initial matter mixing angle when the $\nu_e$ was created. Clearly, for $\cos2\theta_\mathrm{M}^\mathrm{(i)}\sim-1$ and $\cos2\theta\sim1$, almost all of the $\nu_e$s created will be converted to $\nu_\mu$s and $\nu_\tau$s.

The same resonant flavor transformation applies to the $\nu_x$s, so we would therefore expect nearly complete swapping of flavor labels for the $e$ and $x$ flavor neutrinos after encountering adiabatic resonance. While it may seem at first that these effects cancel out and there is no net change in the relative abundances of flavors, recall that neutrinos produced via $e^+ + e^-\rightarrow\nu+\bar{\nu}$ have an overabundance of $\nu_e\bar{\nu}_e$-pairs relative to $\nu_\mu\bar{\nu}_\mu/\nu_\tau\bar{\nu}_\tau$-pairs by a factor of about 5-to-1~\cite{shifuller}. As discussed in the next subsection, this asymmetry at creation will have drastic downstream effects for the final flavor flux ratios after MSW adiabatic flavor transformation.

We have preformed this analysis in the normal mass hierarchy, so $\Bar{\nu}_e$ will not experience this maximal mixing resonance with the other anti-neutrino flavors. In the inverted mass hierarchy $m_1>m_2$, however, we would see maximal flavor mixing between the $\ket{\bar{\nu}_e}$ and $\ket{\bar{\nu}_x}$ states with the $\ket{\nu_e}$ and $\ket{\nu_x}$ states unaffected, the opposite of the case just discussed.

\subsection{Ratio of Flavor Fluxes}

Our simplistic picture of a 2 flavor case can be easily cast into the real 3 flavor case by notable coincidence: the $\ket{\nu_\mu}$ and $\ket{\nu_\tau}$ weak eigenstates are nearly maximally mixed ($\sin\theta_{23}\sim1/\sqrt{2}$~\cite{ParticleDataGroup:2024cfk}). We can therefore rewrite the ``$x$'' flavor as (following Refs.~\cite{BALANTEKIN1999195,PhysRevD.61.123005})
\begin{equation}
    \ket{\nu_x}=\frac{1}{\sqrt{2}}\pr{\ket{\nu_\mu}+\ket{\nu_\tau}},
\end{equation}
and introduce a new flavor $x^\prime$ that is fully decoupled from the $e$ and $x$ flavors:
\begin{equation}
    \ket{\nu_{x^\prime}}=\frac{1}{\sqrt{2}}\pr{\ket{\nu_\mu}-\ket{\nu_\tau}}.
\end{equation}

Recall now that~\citet{shifuller} found that $\nu_e\bar{\nu}_e$ pairs have a ${\sim}$4.7 times greater flux than the $\nu_\mu\bar{\nu}_\mu/\nu_\tau\bar{\nu}_\tau$ pairs for $e^\pm$ annihilation-produced neutrinos in the centers of collapsing SMSs. This dense environment gives the neutrinos created in the $\ket{\nu_e}$ state a higher effective mass than those in the $\ket{\nu_x}$ state on account of the extra CC potential $A_\mathrm{CC}\propto n_e$. These neutrinos stream through the SMS HC almost entirely unimpeded, pass through the MSW resonant flavor mixing density somewhere in the bulk of the HC, and continue to the diffuse outer layers. Since this process is adiabatic for these neutrino energies and SMS density profiles, the net result is an almost complete swapping of flavor labels between $e$ flavor and $x$ flavor neutrinos, the latter representing about one-half of the $\mu$ and $\tau$ flavor neutrinos. Whether this occurs for neutrinos or anti-neutrinos depends on the mass hierarchy. 

In the normal mass hierarchy $m_1<m_2$, the neutrino flavors are maximally transformed. The new ratio of flavors is closer to $\nu_e/\nu_{\mu,\tau}\rightarrow1/5$, an inversion of the original ratio at the SMS center. The other half of $\mu$ and $\tau$ flavor neutrinos, represented by the $x^\prime$ state, are decoupled and hence unaffected. The anti-neutrino flavors are unaffected. This is in contrast to the inverted mass hierarchy $m_1>m_2$, where the MSW resonance inverts the anti-neutrino flavor flux ratio to $\bar{\nu}_e/\bar{\nu}_{\mu,\tau}\rightarrow1/5$, leaving the neutrino flavors unaffected.

Per Figure~\ref{fig:rres}, we predict a wide range of SMS homologous core masses to exhibit nearly complete flavor transformation of core-produced neutrinos or anti-neutrinos at some point in their collapse.  

\section{Neutrino Capture Induced Burning}\label{sec:cap}

For stars burning on the main sequence with low metallicity, the principle $p\text{-}p$ reaction chain follows as~\cite{clayton}
\begin{align*}
    p + p &\rightarrow \ce{^2D} + e^+ + \nu_e, \\
    p + \ce{^2D} &\rightarrow \ce{^3He} + \gamma, \\
    \ce{^3He} +\ce{^3He} &\rightarrow \ce{^4He} + p + p.
\end{align*}
The first step of this chain involves converting a proton into a neutron, a Weak process, making this the primary bottleneck for this reaction.

This bottleneck can be overcome if there is already a supply of free neutrons in the gas. While, normally, free neutrons decay into protons on a rather rather short time scale, neutrons can be made via inverse-$\beta$ decay through the reaction $\Bar{\nu}_e+p\rightarrow n+e^+$. Assuming equal amounts of $\nu_e$ and $\Bar{\nu}_e$, the reverse reaction $\nu_e+n\rightarrow p+e^-$ and neutron decay would basically ensure no such supply can be maintained. 

If, however, the neutrinos experience resonant flavor mixing on their way out of the core, the amount of $\nu_e$ will be heavily suppressed relative to the amount of $\Bar{\nu}_e$.~\citet{Fuller_1997} found that the probability that inverse-$\beta$ decay occurs at radius $r$ is given approximately by
\begin{equation}\label{eq:pdeut}
    P_{\beta^{-1}}\approx4.5\times10^{-6}f_E f_{\Bar{\nu}_e} M_5^\mathrm{HC}\pr{\frac{\left<E_\nu\right>}{1\,\mathrm{MeV}}}\pr{\frac{10^{13}\,\mathrm{cm}}{r}}^2
\end{equation}
where $f_E\approx0.1$ is the amount of gravitational redshift experienced by the neutrinos, $f_{\Bar{\nu}_e}\sim0.4$ is the fraction of all neutrinos which are $\Bar{\nu}_e$, and $M_5^\mathrm{HC}$ is the homologous core mass in units of $10^5\,M_\odot$. Most neutrons created in this process will almost immediately find a proton and fuse to $\ce{^2D}$. If the rate of $n+p\rightarrow \ce{^2D}+\gamma$ is fast compared to both the material ejection and neutron decay rates \emph{and} slow compared to compared to the $p+\ce{^2D}\rightarrow\ce{^3He}+\gamma$ rate, then the probability in Eq.~\eqref{eq:pdeut} is also the fraction of hydrogen converted to $\ce{^2D}$ at a given radius $r$.

\begin{figure}
    \centering
    \includegraphics[width=\linewidth]{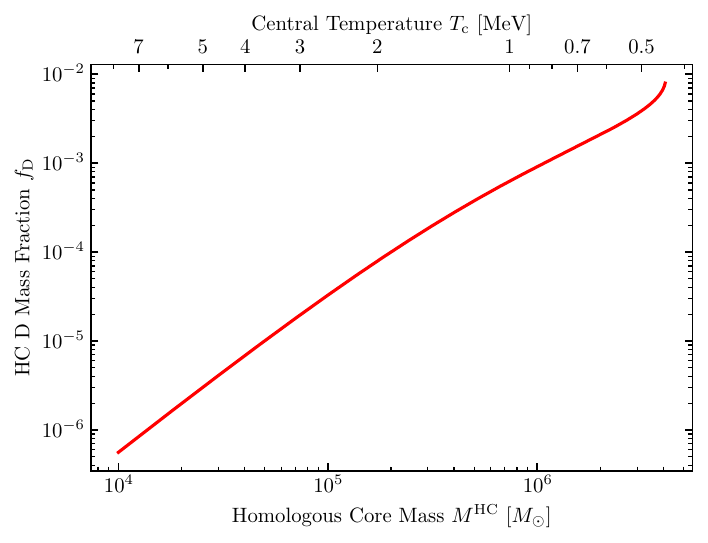}
    \caption{The mass fraction of the SMS homologous core that is converted to deuterium in the region after MSW resonance as a function of the homologous core mass. For $M^\mathrm{HC}\gtrsim\mathrm{few}\times10^5\,M_\odot$, the mass fraction calculation may be dubious (see Figure~\ref{fig:rres}) as the MSW resonance density occurs within the na\"ively calculated gravitational trapped surface horizon at $\rho_\mathrm{c}=\rho_\bullet$.}
    \label{fig:f2D}
\end{figure}

Following this assumption, the mass fraction of the SMS homologous core that is converted to deuterium is
\begin{equation}
    f_{\mathrm{D}} = \frac{1}{M^\mathrm{HC}}\int_{r_\mathrm{res}}^R 4\pi r^2\rho_p(r)P_\mathrm{\beta^{-1}}(r)\,\mathrm{d}r,
\end{equation}
where $\rho_p=X\rho$ is the mass density of free protons if $X$ is the mass fraction of hydrogen. With the standard Lane-Emden coordinate transformation, this mass fraction is
\begin{multline}
    f_{\mathrm{D}}\approx3.3\times10^{-4}f_Ef_{\bar{\nu}_e}\pr{\frac{T_\mathrm{c}}{1\,\mathrm{MeV}}}\\\times \frac{X\rho_\mathrm{c}\pr{10^{13}\,\mathrm{cm}}^2\alpha_\mathrm{LE}}{10^5\,M_\odot}\int_{\xi_\mathrm{res}}^{\xi_3}\varphi^3(\xi)\,\mathrm{d}\xi,
\end{multline}
where $\left<E_\nu\right>\approx1.6\tfrac{F_3(\eta_\nu)}{F_2(\eta_\nu)}T_\mathrm{c}$ is the thermally averaged neutrino energy as a function of the central temperature, $r_\mathrm{res}$ is the location of the MSW resonance density (see the solid red line in Fig.~\ref{fig:rres}), and $\xi_\mathrm{res}=r_\mathrm{res}/\alpha_\mathrm{LE}$. Evaluation of this mass fraction at $\rho_\mathrm{c}=\rho_\bullet$ as a function of SMS homologous core mass is presented in Figure~\ref{fig:f2D}. For a homologous core mass of $M^\mathrm{HC}\sim10^6\,M_\odot$, this means that around a few tenths of a percent of the HC is converted to deuterium, with some unspecified amount more transmuted in the SMS envelope that did not collapse homologously. This makes what is presented in Figure~\ref{fig:f2D} a conservative underestimate of the amount of deuterium produced because $\bar{\nu}_e$s will continue reacting on free protons outside of the homologous core. Depending on local temperatures and densities, the deuterium could then readily react with the surrounding matter, potentially up to $\alpha$ particles. 


\section{Conclusion}

Supermassive stars may yet be a key to understanding the very early universe and the structure formation therein, providing an explanation---however unlikely---to the apparent overabundance of very massive black holes soon after the Big Bang. In this paper, we have discussed their structure, which is well approximated by an $n=3$ polytrope, and how general relativistic corrections to Newtonian gravity cause their eventual rapid collapse. During this infalling period on the short time scale of $\tau_\mathrm{ff}\sim4.8\times10^{-5}(M/10^5\,M_\odot)^{7/4}\,\mathrm{yr}$, the central core will rapidly increase in density and temperature, causing copious creation of $\nu\bar{\nu}$ pairs. Charged and neutral current Weak interactions, as well as the energetics involved in this environment, favor the creation of $\nu_e\bar{\nu}_e$ pairs over $\nu_\mu\bar{\nu}_\mu$ and $\nu_\tau\bar{\nu}_\tau$ pairs by a factor of ${\sim}$5-to-1. We showed that the conditions, at least for homologous core masses of $H^\mathrm{HC}\gtrsim10^4\,M_\odot$, likely do not facilitate appreciable collective oscillation effects as the neutrino number densities are too low. We then proceeded to show that, for SMS homologous core masses between $10^4\,M_\odot < M^\mathrm{HC} \lesssim 10^6\,M_\odot$ and at the central density where a gravitational trapped surface is formed, there exists a density away from the core where there are MSW resonances where flavor conversion is adiabatic. This maximal flavor mixing (at least in the normal mass hierarchy) serves to completely swap the flavor labels between $\nu_e$s with $\nu_\mu$ and $\nu_\tau$, leaving the $\bar{\nu}$s of all flavors unaffected. Therefore, the electron anti-neutrinos will outnumber the electron neutrinos by $n_{\bar{\nu}_e}/n_{\nu_e}\sim2.5$ in the outer layers of the SMS beyond the MSW resonance location. We then show this enhanced $\nu_e \bar{\nu}_e$-asymmetry leads to an appreciable creation of free neutrons in the outer layers of the SMS. Clearly, however, questions for future work abound.

While it is clear that the neutron-to-proton ratio would change in the SMS as a result of these MSW resonance neutrinos, what implications would that have?~\citet{Fuller_1997} showed that there are a wide range of conditions where any free neutrons created would readily find a proton to create deuterium, implying that Eq.~\ref{eq:pdeut} also gives the fraction of hydrogen that that is converted to deuterium. With the Weak bottleneck in the $p\text{-}p$ chain overcome, it would also be reasonable to assume at least a substantial amount of the deuterium would eventually be incorporated into $\alpha$ particles, releasing ${\sim} \ 32\,\mathrm{MeV}$ of energy into the diffuse outer layers of the SMS per reaction. As SMSs are very well described by $n\sim3$ polytropes, there is very little gravitational binding energy holding on to the outer layers. Is this energy deposition enough to perhaps eject material from the already loosely held together SMS? Does the SMS reach a neutrino Eddington-like luminosity from elastic scatterings? That is, can outer layers be ejected by substantial neutrino flux alone? Are these questions made moot by the entirety of the SMS falling into its own black hole regardless? As opposed to our analysis, which tried to be as analytic as possible, all of these questions likely need detailed simulations that includes robust neutrino processes, energy transport, and full general relativity to be answered satisfactorily. Nevertheless, our results suggest that the MSW resonance in SMSs is something worthy of consideration when analyzing these objects, and warrants future study.

Collective flavor oscillation effects were touched upon in this paper, and we determined that their contribution is likely unimportant at least for $M^\mathrm{HC}\gtrsim10^4\,M_\odot$. In this mass range, the number density of neutrinos is such that $\sqrt{2}G_\mathrm{F}n_\nu\sim\Delta m^2/(2E_\nu)\ll\sqrt{2}G_\mathrm{F}n_e$ in the core, i.e.~any collective effects occur on the same timescale as vacuum oscillations and are swamped by matter effects. Collective flavor oscillations would serve to equilibrate the initial flavor asymmetry at the core, severely dampening the MSW enhanced neutron production after MSW resonance. Since collective flavor oscillations are a non-linear effect, however, a full integration of the neutrino flavor quantum kinetic equations in these SMSs is warranted instead of just examining relative energy scales. 

Furthermore, energy scale comparisons suggest that collective effects would be important for the lighter SMS homologous core masses. As the neutrinos are effectively trapped in the core by scatterings for $M^\mathrm{HC}\lesssim\text{few}\times10^5\,M_\odot$ anyway, this regime was not studied in detail in this work, but future work is needed to examine it further. At these higher densities, there may also be substantial $\nu_e$ flux from electron capture on protons, thereby altering the neutron-to-proton ratio in the core and creating an asymmetry in $\nu_e$s and $\bar{\nu}_e$s. The former consideration has implications for energy deposition in the core region during collapse, and the latter has further implications for collective flavor effects as the assumption of $n_{\nu_e}/n_{\bar{\nu}_e}\sim1$ in the core breaks down.

Finally, being such massive objects in the early universe, it is also worth examining dark matter's effect on this entire process. It has been shown that (see, e.g.~\cite{Bisnovatyi-Kogan_1998, mclf,PhysRevD.110.083035,haemmerle,PhysRevD.110.083035}) a substantial dark matter mass fraction in SMSs increases their stability, that is, the critical density at which the SMS is unstable to collapse is increased. This means that, before the gravitational trapped surface is formed, the core is denser and hotter than what's considered in this paper, so the neutrino emissivity during collapse is greatly amplified. This means that more entropy is carried away from the core, causing the homologous core mass to likely be less massive than the figure of $M^\mathrm{HC}\sim0.1M^\mathrm{init.}$ from Ref.~\cite{shifuller}. Another possibility is that dark matter plays a role via dynamical heating, for example in Supermassive Dark Stars~\cite{freese2025earlyformationsupermassiveblack}.

SMSs are of great interest for study, and may hold the key to understanding why there are so many supermassive black holes in the early universe of such high masses. We have examined, being as analytic as possible, the realization that a confluence of neutrino energies and matter densities in SMSs are such that an MSW resonance occurs for the prodigious amount of neutrinos produced in and leaving the core. This effect greatly alters the initial neutrino/anti-neutrino flavor distribution, leading to a substantial amount of free neutrons in the outer layers. Whether this process may provide an observational handle on these objects remains to be seen, but the subsequent evolution of these SMSs post instability will be impacted by this effect. Future work that studies this evolution in more detail with this MSW flavor oscillation resonance will be more complete, offering greater insight into these objects and the early universe in general.

\begin{acknowledgments}
    This work was supported in part by National Science Foundation (NSF) Grants  No.~PHY-2209578 and No.~PHY-2515110 at UCSD and the {\it Network for Neutrinos, Nuclear Astrophysics, and Symmetries} (N3AS) NSF Physics Frontier Center, NSF Grant No.\ PHY-2020275, the Heising-Simons Foundation (2017-228), and the UC San Diego Academic Senate.

    This work made use of the following software packages: \texttt{Jupyter}~\citep{2007CSE.....9c..21P, kluyver2016jupyter}, \texttt{matplotlib}~\citep{Hunter:2007}, \texttt{numpy}~\citep{numpy}, \texttt{python}~\citep{python}, \texttt{scipy}~\citep{2020SciPy-NMeth, scipy_13352243}, and \texttt{Mathematica}~\citep{Mathematica}.  This research has made use of NASA's Astrophysics Data System.  Software citation information aggregated using \texttt{\href{https://www.tomwagg.com/software-citation-station/}{The Software Citation Station}}~\citep{software-citation-station-paper, software-citation-station-zenodo}.
\end{acknowledgments}

\bibliography{bib}

\end{document}